\documentclass[12pt]{article}
\usepackage{graphicx}
\usepackage{amssymb}

\begin{document}
\leftline{\bf CASIMIR FORCES BETWEEN NANOPARTICLES AND SUBSTRATES}
\medskip
\leftline{C. E. Rom\'an-Vel\'azquez$^1$, Cecilia Noguez$^2$, C. Villarreal$^2$ and R. Esquivel-Sirvent$^2$}

\leftline{$^1$Centro de Investigaci\'on en Ciencia Aplicada y Tecnolog\'{\i}a Avanzada, Instituto}
\leftline{Polit\'ecnico Nacional, Av. Legaria 694, Col. Irrigaci\'on, D.F. 11500, M\'exico}  
 \leftline{$^2$Instituto de F\'{\i}sica, Universidad Nacional Aut\'onoma
 de M\'exico, Apartado Postal 20-364,}
\leftline{D.F. 01000,  M\'exico}  

{\bf ABSTRACT}

We study the Casimir force between a nanoparticle and a substrate. We consider the interaction of metal nanoparticles with different substrates within the dipolar approximation. We study the force as a function of the distance for gold and potassium spheres, which are over a substrate of titanium dioxide, sapphire and a perfect conductor. We show that Casimir force is important in systems at the nanometer scale. We study the force as a function of the material properties, radii of the spheres, and the distance between the sphere and the substrate. 

{\bf INTRODUCTION}

Recent advances in micro and nano devices have opened the possibility of studying quantum phenomena that occurs at these length scales. Such is the case of Casimir forces~[1] predicted by the theory of quantum electrodynamics. The textbook example~[2] consist of two parallel neutral conducting plates separated a distance $h$. The plates will attract each other with a force per unit area of roughly one atmosphere when the plates are 35 nm apart.  This force has been measured accurately in different ways. However, a truly parallel plate configuration has been measured only by Bressi {\it et al.}~[3]. The difficulty of keeping the two plates parallel at separations of few nanometers  makes it easier to measure the Casimir force between a large conducting sphere and a plane using microtorsional balances~[4] or atomic force microscopes~[5,6].  In this cases, comparison with the theoretical results obtained for the parallel plates  is done using the proximity theorem~[7]. The proximity theorem is a geometrical approximation that states that if the free energy per unit area $E$ at a given distance between two parallel plates is known, the force between a sphere and a plane is $2 \pi E/R$, where $R$ is the radius of the sphere. This approximation is valid provided the minimum separation between the sphere and the plane is much smaller than $R$. The theorem does not quantify or gives bounds for the ratio between $R$ and the separation with the plane.  Thus, the question improving the theoretical description of the proximity theorem is important from the theoretical and experimental point of view. 

In this work we present a calculation of the Casimir force between a sphere and a conducting plane using a spectral representation approach~[8]. The Casimir force is calculated as a function of  $z$, the minimum separation between the sphere and the plane. The force is studied as a function of the sphere's radius, and the dielectric functions of the sphere, substrate and ambient.

{\bf THEORY}

We consider a system composed by a homogeneous sphere of radio $R$, electrically neutral and dielectric function $\epsilon_s(\omega)$. The sphere is suspended above a substrate which is also electrically neutral and with dielectric function $\epsilon_p(\omega)$. The space above the substrate, where the sphere is immersed, is also electrically neutral and characterized by a dielectric function $\epsilon_a(\omega)$.  We suppose that quantum fluctuations of the electromagnetic field in vacuum induce a charge distribution on the sphere, which origins a dipole moment ${\bf p}_s$. If the sphere is close to a substrate, the charge distribution on  the sphere generates a field which also induces a charge distribution on the substrate, given by a image-dipole moment ${\bf p}_p$. The latter also alters the charge distribution on the sphere through the so-called local-field. The dipole moment in the sphere is affected by the presence of the substrate through the local-field such that
\begin{equation}
 {\bf p}_{s}=  \alpha \left[ {\bf V}^{\rm vac} +  {\bf A} \cdot {\bf p}_ p \right],
\end{equation}
where $\alpha$ is the polarizability of the sphere, $\alpha = R^3 [(\epsilon_s -1) / (\epsilon_s +2)]$, ${\bf V}^{\rm vac}$ is the electromagnetic field in vacuum. Here ${\bf A}$ is the dipole-dipole interaction matrix given by
\begin{equation}
A_{m}^{m'} =  \frac{4 \pi (-1)^m} {(d/R)^3} \frac{2}{3}
 \left[ \frac{1}
{(1+m)! (1-m)! (1+m')! (1-m')!}\right] ^{1/2},
\end{equation}
where $d$ is the distance from the center of the sphere to the substrate, and $m$ and $m'$ indicate the component of the dipole of the sphere and substrate, respectively. Given the symmetry of the system $m$ and $m'$ only has two independent components: one perpendicular ($m=0$) and one parallel ($m=\pm 1$) to the surface of the substrate. Finally, using the image method we can calculate the dipole moment on the substrate, as
\begin{equation}
p_p^{m'} = (-1)^{m'} f_c p_s^{m'}.
\end{equation}
Here $f_c$ is a factor of contrast between the dielectric properties of the ambient and the substrate given by 
\begin{equation}
 f_c = \frac{\epsilon_a - \epsilon_p}{\epsilon_a + \epsilon_p}. 
\end{equation}

Let us define a variable which relates the dielectric properties of the sphere and the ambient as $u= [1 - \epsilon_s/\epsilon_a]^{-1}$, and $x_m = p_s^{m'}/R^{3/2}$ and $g_m = -R^{3/2} V^{\rm vac}_m /4\pi$. Then, we can rewrite each component $m$ of Eq.~(1) as
\begin{equation}
[-u \delta_{mm'} + H_m^{m'}] x_{m'} = g_m ,
\end{equation}
where
\begin{equation}
 H_m^{m'}= n^0 \delta_{mm'} + f_c \frac{R^3}{4 \pi} (-1)^m A_m^{m'},
\end{equation}
with $n^0=1/3$.

We observe that $H_m^{m'}$ is a hermitian matrix if $f_c$ is real, in that case, we can find a unitary matrix {\bf U}, such that, ${\bf U}^{-1} H {\bf U} = 4 \pi n_s$, and using the Green's function, we can obtain a density of states of the variable $u$  for a given direction $m$, as
\begin{equation}
\rho_m (u) = \frac{-1}{\pi} {\rm Im} \left[ \sum_s \frac{(U_{1s}U^{-1}_{s1})_m}{u - n_s} \right] .
\end{equation}
Then, the total density of states is given by $\rho (u) = \sum_m \rho_m (u) = \rho_0 (u) + 2 \rho_1 (u). $

To calculate the force between the substrate and the sphere, we need to consider the quantum fluctuations of the electromagnetic between the isolated sphere, and the sphere near to the substrate. Then, we need to calculate the energy due to the quantum fluctuations~[1] of the system sphere-substrate given by
\begin{equation}
U = \int_0^\infty \frac{\hbar \omega}{2} [ \rho^{sp}(\omega) - \rho^s(\omega)] d \omega,
\end{equation}
where $\rho^{sp}(\omega)$ is the density of the states of the system sphere-substrate, while $\rho^s(\omega)$ is the density of the states of the isolated sphere ($f_c = 0$ or $\epsilon_p = \epsilon_a$). 

The Casimir force is given by $F = - dU/dz$, where $z$ is the distance of separation between the sphere and the substrate, where we have $z=d-R$. Notice that for a given model of the dielectric function of the sphere and the ambient, we can find a relation between $\rho(\omega)$ and $\rho(u)$, as we will show latter in the section of results.

{\bf RESULTS AND DISCUSSION}

In this section we show results for the Casimir force for metal spheres of $10$~nm, $100$~nm and $1000$~nm of radii. The dielectric function of the sphere is given by the Drude model, such that, 
\begin{equation}
\epsilon_s (\omega) = 1 - \frac{\omega_p^2}{\omega (\omega + i/\tau)},
\end{equation}
where $\omega_p$ is the plasma frequency and $\tau$ is the relaxation time of a given material. Then, the density of states as a function of frequency is derived from Eq.~7, and is given by
\begin{equation}
\rho_m(\omega) = \frac{2\omega^2_p}{\pi}\sum_s\sqrt{n_s} \left( U_{1s}U^{-1}_{s1} \right)_m
\left[ \frac{\omega/\tau}{(\omega^2 - n_s \omega^2_p)^2 + (\omega/\tau)^2} \right] .
\end{equation}
We present results for spheres made of potassium (K) with $\hbar \omega_p = 3.8$~eV and $1/\tau=0.105\omega_p$, and gold particles (Au) with $\hbar \omega_p = 8.55$~eV and $1/\tau=0.0126\omega_p$. To achieve the condition of $f_c$ to be real, we have considered substrates which dielectric function is real and constant in a wide range of the electromagnetic spectrum such as TiO$_2$  (titanium dioxide), Al$_3$O$_2$ (sapphire), and a perfect conductor, where $f_c = -0.773, \, -0.516, \,{\rm and}\, -1.0$, respectively. For $\epsilon_p > 1$ then $f_c < 0$ always. In this work, we only consider the case where the sphere is embedded in air, $\epsilon_a=1$. Our theory does not consider retardation effects, therefore we have to work in the limit when $R$ and $z$ is smaller than $c/\omega_p$, with $c$ the speed of light. 

In Fig.~1, we show the energy calculated from Eq.~8 as a function of the separation on units of the radius of the sphere, $z/R$. On left, we show results for spheres of potassium (K) over different substrates. On right, we show results for spheres of gold (Au) over the same substrates. In general, we observe that energy is negative in both cases, and its absolute value is two times larger for gold particles than for potassium particles. We also observe that as absolute value of $f_c$ is larger, then the absolute value of energy is also larger. 

In Fig.~2 we show the force for gold and potassium nanoparticles over a perfect conductor for spheres of radii of $10^1$~nm, $10^2$~nm and $10^3$~nm. In general, we observe that this force is attractive in both cases and is smaller as the radius of the sphere becomes larger at small separations. In particular, at $z=0$~nm the force for a sphere with $R=10$~nm is ten times more attractive than the one with $R=10^2$~nm, and it is a hundred of times more attractive than the force of the sphere with $R=10^3$~nm. We also observe that as the radius of the sphere increases the force decreases slowly as function of $z$. Then, the force for the larger sphere with $R=10^3$~nm is almost flat as the separation increases from 0 to 40~nm. On the other hand, for a sphere of radius $R=10$~nm the force decreases very fast as a function of $z$, in this case the force decreases about three order of magnitude as the separation of the sphere goes from 0 to 40~nm.

In Fig.~3 we compare the force between nanoparticles with radii of $R=10$~nm and made of different materials (potassium and gold), over the same substrate. On left we show results for a sapphire substrate, while on right we show results for a perfect conductor substrate. As we obtained before, the force is more attractive for the perfect conductor substrate than the sapphire one. For both substrate, we observe that the force between gold particles and the substrate is more attractive than the force with potassium particles. Note that gold particles have a plasma frequency more than two times larger than the plasma frequency of potassium particles.

In Fig.~4, we show as a function of energy (eV), the difference of the density of state between the sphere near a substrate and the isolated sphere, $\rho(\omega)=\rho(\omega)^{\rm sp} - \rho(\omega)^{\rm s}$. We show results for potassium (upper plots) and gold nanoparticles (lower plots) over  a substrate of sapphire (left) and a perfect conductor (right). Different curves correspond to different separations of the sphere to the substrate on units of the radius of the nanoparticle, $z/R$, from 0.0 to 4.0. We found that for gold particles over a perfect conductor, $\rho$ shows three different peaks, two positive and one negative, when the sphere is touching the substrate ($z/R = 0$). At $z/R=0$, the positive peak at lower energies (at about 4.25~eV) corresponds to electromagnetic modes (EM) on the sphere which are perpendicular to the substrate plane, while the positive peak at about 4.6~eV corresponds to EM modes on the sphere which are parallel to the substrate plane. The latter peak is about two times larger than the one corresponding to perpendicular modes. At larger energies we found a negative peak that corresponds to the EM modes of the isolated sphere. As the separation $z/R$ increases, the strength of all the peaks decreases and the positive peaks are blue-shifted and are overlap, while the negative peak does not move its position respect to the energy. For large separations, the density of states is almost null. The same is observed for Au/Al$_2$O$_3$, however the positive peaks are at larger energies and are overlap between them. On the other hand, we found that K particles show a smaller density of states than Au particles, and $\rho$ only shows two peaks, one negative and one positive. The positive peak corresponds to all the EM modes (perpendicular and parallel to the substrate plane) which are all at the same energy, while the negative peak also corresponds to the EM modes of the isolated sphere. Notice than these peaks are wider than those show for Au particles, this fact is due to the difference in $\tau$ of K and Au, and to the fact that perpendicular and parallel EM modes can not be distinguished giving rise to a wide structure.

This work has been partly financed by CONACyT grant No. 36651-E and by DGAPA-UNAM grant No. IN104201. 

{\bf REFERENCES}
\medskip

\small

\leftline{[1] H. B. G. Casimir, Proc. Kon. Ned. Akad. Wet. {\bf 51}, 793 (1948). }
\leftline{[2] P. W. Milonni, {\it The quantum vacuum, an introduction to quantum electrodynamics},
Academic}
\leftline{Press Inc, San Diego (1994).} 
\leftline{[3] G. Bressi, G. Carugno, R. Onofrio, and G. Ruoso, Phys. Rev. Lett., {\bf 88}, 041804-1 (2002).}
\leftline{[4] S. K. Lamoreaux, Phys. Rev. Lett. {\bf 78}, 5 (1997).}
\leftline{[5] H. B. Chan, V. A. Aksyuk, R. N. Kliman, D. J. Bishop and F. Capasso, Science {\bf 291}, 1942 (2001).}
\leftline{[6] U. Mohideen and Anushree Roy, Phys. Rev. Lett. {\bf 81}, 4549  (1998). }
\leftline{[7] B. W. Harris, F. Chen, and U. Mohideen, Phys. Rev. A {\bf 62}, 052109 (2000).} \leftline{[8]  C. E. Rom\'an, C. Noguez, R. G. Barrera, Phys. Rev. B, {\bf 61}, 10427  (2000).} 
\newpage
\begin{figure}[h]
\centerline {
\includegraphics[width=6.5in]{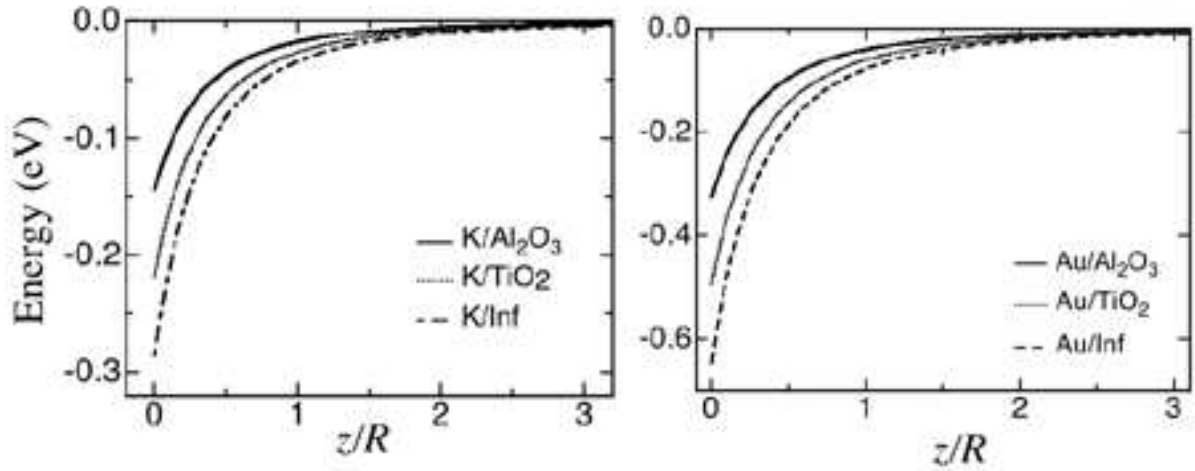}
}
\caption{Energy as a function of the separation on units of radius ($z/R$) for K and Au spheres over substrates of Al$_2$O$_3$, TiO$_2$, and a perfect conductor (Inf).}
\end{figure}

\begin{figure}[h]
\centerline {
\includegraphics[width=6.5in]{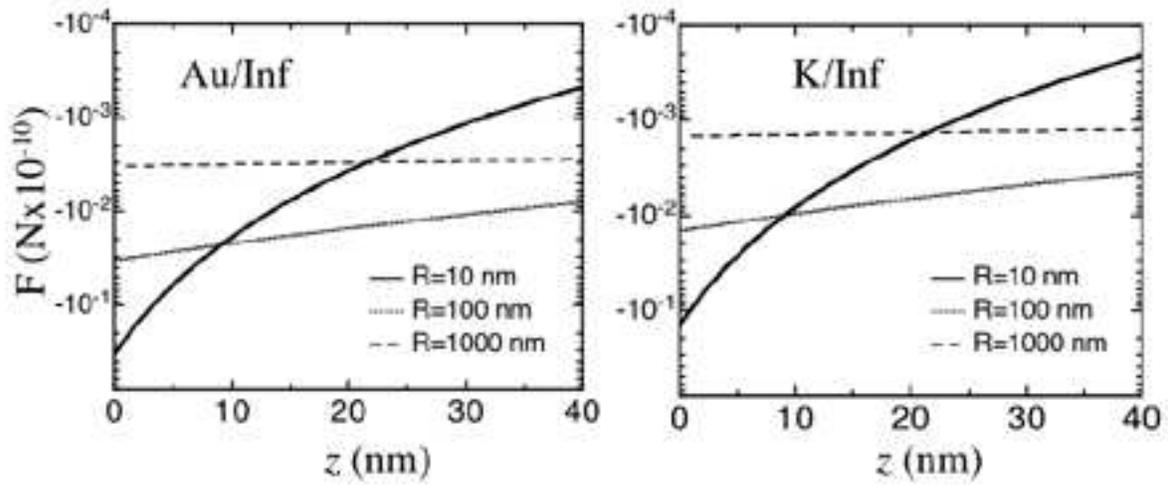}
}
\caption{Force as a function of the separation for K and Au spheres over a perfect conductor (Inf) substrate and spheres of different radii.}
\end{figure}

\begin{figure}[h]
\centerline {
\includegraphics[width=6.5in]{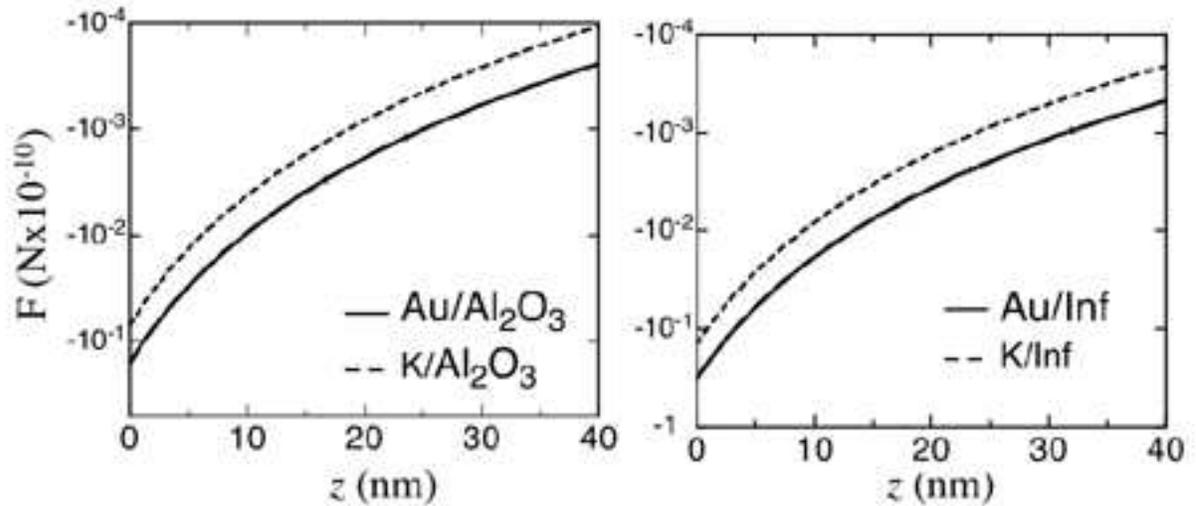}
}
\caption{Force as a function of the separation for K and Au spheres over Al$_2$O$_3$ (left) and perfect conductor (right) substrates. Spheres with radii of $R=10$~nm.}
 \end{figure}

\begin{figure}[h]
\centerline {
\includegraphics[width=6.5in]{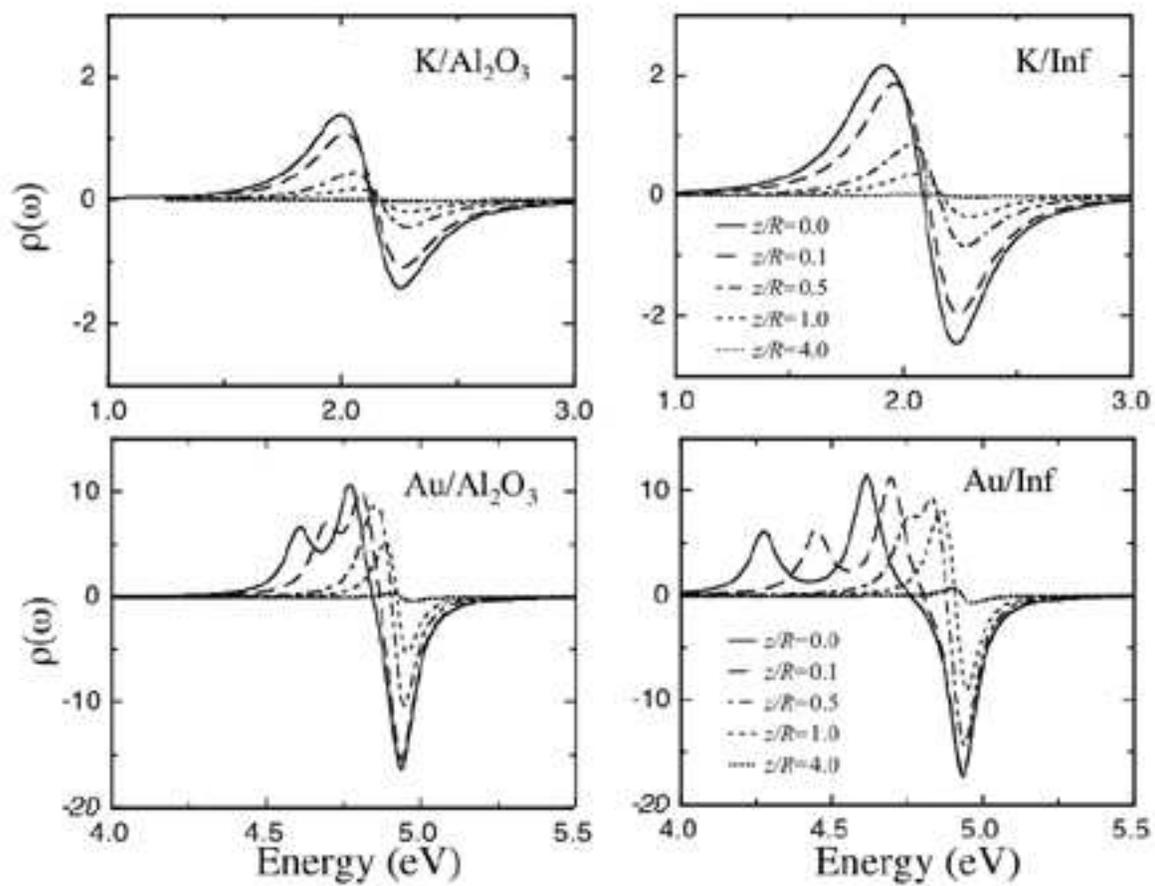}
}
\caption{\small $\rho$ for Au and K nanoparticles over sapphire and a perfect conductor.}
 \end{figure}

 \end{document}